\documentstyle[aps,prl,twocolumn]{revtex}
\newcommand{\od}[2]{\frac{d #1}{d #2}}
\newcommand{\vu}{{\bf u}}
\newcommand{\vp}{{\bf p}}
\newcommand{\be}{\begin{equation}}
\newcommand{\ee}{\end{equation}}
\newcommand{\ba}{\begin{eqnarray}}
\newcommand{\ea}{\end{eqnarray}}
\begin{document}
\draft
\title{
    The Mikheyev-Smirnov-Wolfenstein effect as a probe of 
                     the solar interior
}
\author{
         L. H. Li$^1$, Q. L. Cheng$^1$, Q. H. Peng$^2$ and H. Q. Zhang$^1$
}

\address{
         $^1$ Purple Mountain Observatory, Academia Sinica,
         Nanjing 210008, PR China
}
\address{
         $^2$ Department of Astronomy, Nanjing University,
         Nanjing 210008, PR China
}

\maketitle

\begin{abstract}
We relate the MSW effect to the efective absorption of the electronic
collective motion energy by retaining the imaginary part of the index of
refraction associated with the charged-current scattering and show that
the small angle MSW solution to the solar neutrino anomaly can be used
as a probe of the physical conditions of the solar interior if it is
correct. We find that the constraint on the absorption imposed by the
small angle MSW solution and the theoretical estimate of the absorption
by the Boltzmann kinetic theory are consistent, which shows that a
consistent theoretical picture can be developed when plasma absorption
processes are taken into account.
\end{abstract}


If, based on its apparent success, the small angle MSW
solution\cite{MSW-solution-1,MSW-solution-2,MSW-solution-3}
is accept as the solution for the solar neutrino
problem\cite{SSM,Homestake-1,Kamiokande-2,Super-Kamiokande-3,%
SAGE-4,GALLEX-5}, it can be used as a probe of physical conditions in
the solar interior. Since the MSW effect\cite{MSW} can be attributed to
the resonant scattering of neutrinos off electrons in the solar
interior, it must be sensitive to the absorption of not only the
neutrino energy but also the electron energy. The effective absorption
can be represented in terms of complex indices of refraction. In order
to take into account the absorption, we rewrite the index of refraction
$n_e$ associated with the charged-current scattering in such a
suggestive form\cite{MSW}:
\begin{equation}
  n_e=1+\frac{G_F}{4\pi E\alpha\lambda_c}
   \omega_e^2, \label{index}
\end{equation}
where $G_F$, $\alpha=e^2$, $\lambda_c=m_e^{-1}$, $E$ and $\omega_e$ are,
respectively, the Fermi constant, fine structure constant, electron
Compton wavelength, neutrino energy and electronic plasmon energy,
noticing that we have set $\hbar=c=1$. Since the MSW effect depends on
an effective density-dependent contribution to the neutrino
mass, longitudinal plasmons or Langmuir waves due to electrostatic
oscillations of free electrons in the solar plasma make
sense\cite{ABR-84,Swanson-89,Goldston-Rutherford-95}:
\be
  \omega_e=(\omega_{pe}^2+3V_{Te}^2k'^2)^{1/2}+i\Gamma',  \label{e-state}
\ee
where $\omega_{pe}=(4\pi N_ee^2/m_e)^{1/2}$, $V_{Te}=(T/m_e)^{1/2}$, $k'$
and $\Gamma'$ are, respectively, the electronic plasma frequency,
characteristic thermal velocity of electrons, plasmon momentum (or
wavenumber) and effective absorption coefficient of the plasmon.
The fact that $\omega_e$ is complex means that the plasmon or the
coherent motion of electrons in the plasma $e^{-i\omega_e t}$ will
decay. If the plasma is not uniform, all dependent variables will depend
on the location. Obviously, if $k'=0$ and $\Gamma'=0$, then
$\omega_e=\omega_{pe}$, and hence $E(n_e-1)=G_FN_e$, as assumed in the
MSW theory\cite{MSW,MSW-solution-1,MSW-solution-2,MSW-solution-3}. The
purpose of this letter is to investigate how the small angle MSW
solution constrains $\Gamma'$ in the solar plasma.

For simplicity, we consider two neutrino flavors. If we use the flavor
eigenstates $|\nu_e\rangle$ and $|\nu_\mu\rangle$ as the basis,
the time evolution of the neutrino state vector
$
  |\nu(x)\rangle=a_e(x)|\nu_e\rangle+a_\mu(x)|\nu_\mu\rangle
$
in matter in the relativistic limit is governed by the equation
\cite{MSW,MSW-solution-3}
\be
 i\od{}{x}\left(\begin{array}{c} a_e \\ a_\mu  \end{array}\right) 
    =  \frac{\Delta_V}{2}
\left(\begin{array}{cc} 
   M(x)  & \sin2\theta_V \\ 
   \sin2\theta_V    & -M(x)
\end{array}\right)
  \left(\begin{array}{c} a_e \\ a_\mu
  \end{array}\right) \label{damp}
\ee
where $x=ct$, $M(x)=\sqrt{2}E(n_e(x)-1)/\Delta_V-\cos2\theta_V$,
$\delta m^2=m_2^2-m_1^2$, $\Delta_V=\delta m^2/2E$, $\theta_V$ 
is the vacuum mixing angle. $M_r(x)=\mbox{Re}\,M(x)$ and
$M_i(x)=\mbox{Im}\,M(x)$ can be approximately cast as follows:
\begin{mathletters}
\ba
  M_r(x) & \approx & \sqrt{2}G_FN_e(x)(1+3k^2-\Gamma(x)^2)/\Delta_V
              \nonumber \\
         & &  -\cos2\theta_V,  \label{mr} \\
  M_i(x) & \approx & 2\sqrt{2}G_FN_e(x)\Gamma(x)/\Delta_V, \label{mi}
\ea
\end{mathletters}
where $k=k'/k_{D}$, $\Gamma(x)=\Gamma'(x)/\omega_{pe}(x)$, and
$k_D=\omega_{pe}/V_{Te}$. We have assumed $k^2\ll 1$, which is
reasonable because when $k\gtrsim 0.3$ the Landau
damping\cite{ABR-84,Swanson-89,Goldston-Rutherford-95,Brodin-97} will
cut in. The fact that $M(x)$ is complex implies that there is no
$N_e(x_c)$ so that $M(x_c)=0$ no matter $N_e(x)$ is large or small.
Novertheless, if the absorption $\Gamma$ is small enough, the level
crossing still occurs when  
\be
  M_r(x_c)\ge 0.
\ee
Obviously, $k$ favors resonance, while $\Gamma$ disfavors resonance as
it should.

At a free electron number density, $N_e(x)$, the light (L) and heavy (H)
local mass eigenstates\cite{Haxton-95,MSW-solution-3} are
\begin{mathletters}
\begin{eqnarray}
  |\nu_L(x)\rangle &=& \cos\theta(x)|\nu_e\rangle
    -\sin\theta(x)|\nu_\mu\rangle \label{lem} \\
  |\nu_H(x)\rangle &=& \sin\theta(x)|\nu_e\rangle
    +\cos\theta(x)|\nu_\mu\rangle, \label{hem}
\end{eqnarray}
\end{mathletters}
which have complex eigenvalues $\pm\case{1}{2}\Delta(x)$, where
\be
  \Delta(x) = \Delta_V[M(x)^2+\sin^22\theta_V]^{1/2}, \label{dlt}
\ee
and $\theta(x)$ satisfies
\begin{mathletters}
\ba
  \sin2\theta(x) &=& \frac{\sin2\theta_V}{\Delta(x)/\Delta_V}, \label{sin} \\
  \cos2\theta(x) &=& \frac{-M(x)}{\Delta(x)/\Delta_V}. \label{cos}
\ea
\end{mathletters}
Since
\be
  \Delta(x_c)=
\Delta_V(\sin^22\theta_V-4\Gamma_c^2\cos^22\theta_V)^{1/2},
\ee
the absorption will affect the width of the resonance: $2\Delta(x_c)$,
where $\Gamma_c=\Gamma(x_c)$. The corresponding resonance
distance\cite{MSW-solution-3} is
\be
  \delta x=2\left[-\frac{\dot{N}_e}{N_e}\right]^{-1}_{x_c}
    (\tan^22\theta_V-4\Gamma^2)^{1/2}, \label{dx}
\ee
where $\dot{N}_e(x)=dN_e(x)/dx$ is the electronic density gradient.
Because the small angle MSW solution gives the best fit to existing
solar neutrino data, $\sin^22\theta_V\approx 8\times 10^{-3}$, and
$\delta x\ge 0$, we obtain
\be
  |\Gamma_c|\le\case{1}{2}\tan2\theta_V\approx 0.04,
\ee
which is the upper limit of $|\Gamma_c|$ in the solar interior where the
resonance takes place.

The fact that the eigenvalues are complex implies that the neutrino
states will decay, too. In order to see this we express the neutrino
state vector $|\nu(x)\rangle$ in the matter in terms of the local mass
eigenstates
$
  |\nu(x)\rangle=a_H(x)|\nu_H(x)\rangle+a_L(x)|\nu_L(x)\rangle.
$
Consequently, the evolution equation becomes
\be
 i\od{}{x}\left(\begin{array}{c} a_H \\ a_L  \end{array}\right) 
    =\left(\begin{array}{cc} \case{1}{2}\Delta(x)   & i\alpha(x) \\
          -i\alpha(x)  & -\case{1}{2}\Delta(x) \end{array}\right)
  \left(\begin{array}{c} a_H \\ a_L
  \end{array}\right), \label{diagonal}
\ee
where
\[
  \alpha(x)\approx\frac{\Delta_V}{2}
    \frac{\sqrt{2}G_F\dot{N}_e(x)\sin2\theta_V}{\Delta(x)^2}
\]
enforces mixing of the mass eigenstates governed by the density
gradient, where we have assumed $d{\Gamma}/dx=0$. When the off-diagonal
elements can be neglected with respect to the diagonal elements, i.e.,
when
\begin{eqnarray*}
  \gamma(x) & = & \left|\frac{\Delta(x)}{\alpha(x)}\right| \nonumber \\
            & \approx & \frac{\sin^22\theta_V}{\cos2\theta_V}
                  \frac{|\Delta(x)^3|}{\Delta_V^2\sin^32\theta_V}
                  \left|\frac{\dot{N}_e(x)}{N_c}\right|^{-1}\gg 1,
\end{eqnarray*}
the neutrino will propagate adiabatically through the matter. The
absorption tends to weaken this condition, for example, near the
crossing point, it reads
\be
  \gamma_c= \gamma^0_c
          (1-4\Gamma_c^2\cot^22\theta_V)^{3/2}<\gamma^0_c.  \label{gmc}
\ee
where $\gamma^0_c=\gamma(x_c; \Gamma_c=0)$. Since the detected solar
$^7$Be neutrino flux\cite{Homestake-1,Kamiokande-2,Super-Kamiokande-3,%
SAGE-4,GALLEX-5} is much less than the predicted\cite{SSM}, the $^7$Be
neutrinos must propagate adiabatically in the sun, the absorption
correction factor thus should not depart from unity significantly, which
demands
\be
  |\Gamma_c|\ll 0.04.  \label{gmll04}
\ee
Of cause, the absorption also affects the adiabatic
boundary\cite{Haxton-95} of $\gamma_c\sim 1$ in the
$\delta m^2/E-\sin^22\theta_V$ plane.

When $\gamma_c\gg 1$, the adiabatic states are
\begin{mathletters}
\ba
  |L\rangle &=& e^{-ia(x_0;\,x)+b(x_0;\,x)}|\nu_L(x)\rangle, \\
  |H\rangle &=& e^{ia(x_0;\,x)-b(x_0;\,x)}|\nu_H(x)\rangle,
\ea
\end{mathletters}
where
\begin{mathletters}
\ba
  a(x_0;\,x) &=& \case{1}{2}\int_{x_0}^x\Delta_r(x')dx', \\
  b(x_0;\,x) &=& \case{1}{2}\int_{x_0}^x\Delta_i(x')dx', \label{bx}
\ea
\end{mathletters}
where $x_0$ is the production point of the basis states,
$\Delta_r(x)=\mbox{Re}\,\Delta(x)$, $\Delta_i(x)=\mbox{Im}\,\Delta(x)$.
Obviously, they  will decay provided that $b(x)\ne 0$. If $b(x)>0$, the
local heavy mass eigenstate will continuously jump down into the local
light mass eigenstate, while the local light mass eigenstate will
continuously jump 
up into the local heavy mass eigenstate; if $b(x)<0$ the processes are
reversed. Expressions~(\ref{dlt}), (\ref{cos}), (\ref{bx}) and
condition~(\ref{gmll04}) show that $b(x)<0$ if both the plasmons are indeed
absorbed ($\Gamma<0$) and the plasma density is supercritical
($M_r>0$) or both the plasmons are excited ($\Gamma>0$) and the plasma
density is subcritical ($M_r<0$). However, it is still convenient to use
these states as the basis states in the region for which there are no
transitions\cite{MSW-solution-3}. Under the adiabatic approximation, we
have $|\nu(x)\rangle=a_1|L\rangle+a_2|H\rangle$ in which
the linear combination coefficients $a_i$ (i=1, 2) are determined by the
initial condition, hence the average probability of detecting an
electron neutrino at the Earth with the initial condition
$|\nu(x_0)\rangle=|\nu_e\rangle$ is
\ba
  P^{ad}_{\nu_e}=\case{1}{2}(1+\cos2\theta_0\cos2\theta_V)\cosh2b_V 
           \nonumber\\
    +\case{1}{2}(\cos2\theta_0+\cos2\theta_V)\sinh2b_V,  \label{pe}
\ea
where $\theta_0=\theta(x_0;\,\Gamma=0)$, $b_V=b(x_0;\,R_\odot)$. This
expression will reduce to the well-known adiabatic probability if
$\Gamma\equiv 0$ or $b_V=0$, noticing that $\sinh2b_V=0$ and
$\cosh2b_V=1$ when $b_V=0$. This shows that the absorption also affects
the detection probability.

However, it is necessary to go beyond the adiabatic approximation
because the solar pp and $^8$B neutrinos are suppressed
partially\cite{MSW-solution-1,MSW-solution-2,MSW-solution-3,SSM,%
Homestake-1,Kamiokande-2,Super-Kamiokande-3,SAGE-4,GALLEX-5}. We follow
Parke\cite{MSW-solution-3} to do so. Eqs.~(\ref{dx}), (\ref{gmc}) and
(\ref{pe}) show that the critical region for nonadiabatic behavior
occurs in a narrow region $[x_-,\ x_+]$ (for small $\theta_V$)
surrounding the crossing point $x_c$ and that this behavior is
controlled by both the density gradient $[\dot{N}_e/N_e]_{x_c}$ and the
absorption $\Gamma_c$ at the crossing point, where
$x_\pm=x_c\pm\case{1}{2}\delta x$. See Fig.~1 of
Bethe\cite{MSW-solution-3} for graphical understanding. When an
initial electron neutrino at $x_0$ approaches to the inner boundary of
the nonadiabatic region, its state becomes
\ba
  |\nu(x_-)\rangle &=&\cos\theta_0e^{-ia_-+b_-}|\nu_L(x_-)\rangle 
     \nonumber\\
  & & +\sin\theta_0e^{ia_--b_-}|\nu_H(x_-)\rangle,
\ea
where $a_-=a(x_0;\,x_-)$, $b_-=b(x_0;\,x_-)$. As the neutrino goes
through the nonadiabatic region to reach its outer boundary $x_+$, we
have the mixed states according to Parke\cite{MSW-solution-3}
\begin{mathletters}
\ba
  |\nu_L(x_-)\rangle &\rightarrow& a_1|\nu_L(x_+)\rangle
      +a_2|\nu_H(x_+)\rangle, \\
  |\nu_H(x_-)\rangle &\rightarrow& -a_2^*|\nu_L(x_+)\rangle
      +a_1^*|\nu_H(x_+)\rangle,
\ea
\end{mathletters}
where $a_i$ (i=1, 2) are determined by the nature of the transition
point and satisfy $|a_1|^2+|a_2|^2=1$. From the upper boundary on, the
eigenstates $|\nu_L(x)\rangle$ and $|\nu_H(x)\rangle$ will evolve
adiabatically starting with $|\nu_L(x_+)\rangle$ and
$|\nu_H(x_+)\rangle$ respectively. Therefore, the state vector of the
neutrino reads as follows:
\be
  |\nu(x)\rangle=A|\nu_L(x)\rangle+B|\nu_H(x)\rangle, \label{final}
\ee
in the detection region $x\ge x_c$, where
\begin{mathletters}
\ba
  A&=&a_1\cos\theta_0e^{-iA_++B_+}
    -a_2^*\sin\theta_0e^{-iA_-+B_-} \\
  B&=&a_2\cos\theta_0e^{iA_--B_-}
    +a_1^*\sin\theta_0e^{iA_+-B_+}
\ea
\end{mathletters}
in which $A_\pm=a_+\pm a_-$, $B_\pm=b_+\pm b_-$, $a_+=a(x_+;\,x)$ and
$b_+=b(x_+;\,x)$. 

Substituting Eqs.~(\ref{lem}) and (\ref{hem}) into
Eq.~(\ref{final}), one can find the amplitude for producing, in the
solar core $x_0$, and detecting, on the Earth, an electron neutrino
after passage through resonance: $A_e=A\cos\theta_V+B\sin\theta_V$. Thus
the probability of detecting this neutrino as an electron neutrino after
averaging over both the production and the detection positions is given
by
\ba
  P_{\nu_e}&=&|a_1|^2(\cos^2\theta_0\cos^2\theta_Ve^{2B_+}
    +\sin^2\theta_0\sin^2\theta_Ve^{-2B_+}) \nonumber \\
           &+ &|a_2|^2(\cos^2\theta_0\sin^2\theta_Ve^{2B_-}
    +\sin^2\theta_0\cos^2\theta_Ve^{-2B_-}),  \nonumber \\
      \label{general}
\ea
where $b_+=b(x_+;\,R_\odot)$. If $B_\pm=0$, we reproduce the well-known
Parke formula\cite{MSW-solution-1}:
\be
  P^{Parke}_{\nu_e}=\case{1}{2}+(\case{1}{2}
    -P_x)\cos2\theta_0\cos2\theta_V, \label{parke}
\ee
where $|a_1|^2=1-|a_2|^2$ and $P_x=|a_2|^2$ is the probability of
transition from $\nu_H(x_-)\rangle$ to $\nu_L(x_+)\rangle$ (or vice
versa) as the neutrino goes through the nonadiabatic region.
Parke\cite{MSW-solution-3} and Haxton\cite{Haxton-95} have worked out
$P_x$ by using a linear density profile, it is natural to generalize it
to our case with $\Gamma\ne 0$:
\be
  P_x=\exp({-\frac{\pi}{2}\gamma_c}).
\ee

Because the small angle MSW solution with $B_\pm=0$ gives the best fit
to existing solar neutrino data, the solar neutrino experiments
require $|B_\pm|\ll 1$. Using the mean density $\rho_\odot\sim 1$
g cm$^{-3}$, the mean number of electrons per nucleon $Y_e=\case{1}{2}$,
we can estimate $|B_\pm|\sim 10^2 |\overline{\Gamma}|$, where
$\overline{\Gamma}$ is the mean absorption coefficient in the solar
interior. Consequently, the small angle MSW solution demands
\be
  |\overline{\Gamma}|\ll 10^{-2}.
\ee
If we assume that the resonance region is near the production region,
then $b_-\approx 0$, hence $B_+\approx B_-\approx b_+$. Since 
$\Delta_i(x)\approx -\Gamma(x)2\sqrt{2}G_FN_e(x)\cos2\theta'(x)$, where
$\theta'(x)=\theta(x;\, \Gamma=0)$, and $\cos2\theta'(x)>0$ when
$x>x_+$, then $\Delta_i(x)>0$ when $\Gamma(x)<0$. Therefore, $B_\pm>0$
when $\Gamma(x)<0$, but $B_\pm<0$ when $\Gamma(x)>0$. Since numerical
evaluation shows that Eqs.~(\ref{general}) and (\ref{parke}) give the
same results when $B_\pm\lesssim 10^{-3}$, we may infer that
$\Gamma(x>x_c)<0$ and $|\overline{\Gamma}|\lesssim 10^{-5}$ in the solar
interior.

The Boltzmann kinetic theory formulated in the Boltzmann
equation\cite{ABR-84} can be used to estimate the effective absorption
coefficient of the electronic energy. The effective energy transition
rate $Z_{ei}$ and momentum transition rate ${\bf R}_{ei}$ from electrons
to ions per electron in the elastic scattering process
$\vp_e+\vp_i\rightleftharpoons \vp'_e+\vp'_i$ (Chapter~3 of
\cite{ABR-84}) can be estimated through Tayler-expanding these rates
near local thermodynamic equilibrium between electrons and ions as
follows 
\begin{mathletters}
\ba
  Z_{ei}&=&-\nu_Z^{ei}(T_e-T_i), \\
  {\bf R}_{ei}&=&-\nu_R^{ei}(T_e-T_i)(m_e\vu_e-m_i\vu_i),
\ea
\end{mathletters}
where $\nu_Z^{ei}$ and $\Gamma'=-\nu_R^{ei}(T_e-T_i)$ are the effective
energy and momentum transition (or absorption) coefficients of electrons
in the plasma, $\vu_{e(i)}$ is the coherent motion velocity of electrons
(ions), noticing that $Z^{eq}_{ei}=0$ and ${\bf R}^{eq}_{ei}=0$ when
electrons and ions have the same temperature $T_e=T_i$. Since the
effective energy and momentum transition rates are dependent on each
other, $\Gamma'=-\nu_Z^{ei}(T_e-T_i)/T_e$. Therefore, we need only the
effective energy transition coefficient $\nu_Z^{ei}$, which may be
estimated by the thermal timescale of the sun, the timescale that the
stored thermal energy of the sun is used up via radiation at the present
radiation power. The thermal timescale is equal to the gravitational
timescale, which is the ratio of the gravitational energy to the total
luminosity\cite{SSM} 
\be
  t_{gravity}\sim GM_\odot^2/R_\odot L_\odot\approx 10^7\,\mbox{yr}.
\ee
So we may have
\be
  \Gamma'\sim -10^{-14}(T_e-T_i)/T_\odot\, \mbox{sec}^{-1}.
\ee
Since $\omega_{pe}\gg 1$ sec$^{-1}$ and $T_e\gtrsim T_i$ in the solar
interior, we know $\Gamma=\Gamma'/\omega_{pe}$ is negtive and
$|\Gamma|\ll 10^{-5}$ in the solar interior. This shows that the small
angle MSW solution to the solar neutrino problem and the Boltzmann
kinetic theory are consistent. A treatment of the plasma processes
including absorption thus verifies that a consistent theoretical picture
can be developed.

This research belonged to project 19675064 supported by NSFC and was also
supported in part by CAS.

\end{document}